\documentclass[twocolumn,usenatbib,amsmath,amssymb,amsbsy]{mn2e}

\usepackage{natbib}
\usepackage{amsbsy}
\usepackage{amssymb}
\usepackage{amsmath}

\usepackage{graphicx}

\begin{document}

\title[Pulsar state switching, timing noise and free precession] {Pulsar state switching, timing noise and free precession}

\author[Jones]{D.I. Jones$^1$\\
  \\
  $^1$ School of Mathematics, University of Southampton, Southampton
  SO17 1BJ, UK \\
 }

\maketitle

\begin{abstract}
Recent radio pulsar observations have shown that a number of pulsars display interesting long term periodicities in their spin-down rates.  At least some of these pulsars also undergo sharp changes in pulse profile.  This has been convincingly attributed to the stars abruptly switching between two different magnetospheric states.  The sharpness of these transitions has been taken as evidence against free precession as the mechanism behind the long term variations.  We argue that such  a conclusion is premature.  By performing a simple best-fit analysis to the data, we show that the relationship between the observed spin and modulation periods is of approximately the correct form to be accounted for by the free precession of a population of neutron stars with strained crusts, the level of strain being similar in all of the stars, and consistent with the star retaining a memory of a former faster rotation rate.  We also provide an argument as to why abrupt magnetospheric changes can occur in precessing stars, and how such changes would serve to magnify the effect of precession in the timing data, making the observation of the precession more likely in those stars where such switching occurs.  We describe how future observations could further test the precession hypothesis advanced here.

 \end{abstract}


%%%%%%%%%%%%%%%%%%%%%%%%%%%%%%%%%%%%%%
\section{Introduction} \label{sect:intro}
%%%%%%%%%%%%%%%%%%%%%%%%%%%%%%%%%%%%%%

Radio pulsars are remarkably stable rotators.  However, as was apparent from soon after the discovery of the first pulsar, in addition to a smooth secular spin-down, they exhibit other interesting timing features, including sudden spin-ups known as glitches, periodic variations in spin-down rate possibly attributable to free pression, and low-level random timing fluctuations known as timing noise.  

The study of all these forms of spin-down variation are of great interest as probes of neutron star structure.  The glitch and precession phenomena are believed to depend sensitively on the coupling between the solid and superfluid phases of the star (see e.g. \citet{lel99}), while the origin of the timing noise is less well understood, and may be caused by process internal to the star or external (i.e. in the pulsar magnetosphere \citep{cg81}).   The reported existence of sustained long-period free precession is of particular interest, as  such motions are inconsistent with the pinned superfluid model that has become the standard explanation of large glitches \citep{ja01, le01,  link03, link06}.  Additionally, timing noise is expected to limit the performance of pulsar timing arrays, whose aim is to detect low frequency (nano-Hertz) gravitational waves \citep{jetal05}.  As pointed out recently, if timing noise could be modelled and removed from the timing solution, the sensitivity of pulsar timing arrays could be improved \citep{letal10}.  There are therefore both theoretical and practical motivations for understanding better the origin of timing noise.

As summarised below, recent radio observations have shown that some pulsars with harmonic features in their timing residuals undergo sudden sharp changes in pulse profile which, in one case at least, correlates with sharp changes in spin-down torque \citep{ketal06, letal10}.  As was convincingly argued, this is good evidence for the star switching back and forth between two different magnetospheric states.  The sharpness of this switching has been taken as evidence that free precession is not a viable mechanism to explain the harmonic timing data \citep{letal10}, thereby removing the theoretical difficulty discussed above.

However, as we argue below, we believe it is premature to abandon the precession hypothesis as an explanation for the harmonic timing data.  We give two main arguments to support this.  Firstly, the ratios $P/P_{\rm mod}$ of the spin periods $P$ (of order a second) to the long term periodicities $P_{\rm mod}$ (of order a few years)  are of the right magnitude to be accounted for by free precession of a star whose crustal stress is due to it retaining a memory of a former faster rotation rate;  few (if any) other known processes act on such long timescales.  We find that the implied breaking strains must be at least as large as $5 \times 10^{-4}$, with minimum birth spin frequencies of $50$ Hz if only the crust participates in the precession; these numbers increase to $5 \times 10^{-2}$ and $500$ Hz if the whole star precesses as one.  That free precession of a strained star, with these sorts of minimum strains and birth frequencies, is a viable explanation for the timing behaviour of PSR B1828-11 has already been demonstrated in detail by \citet{cul03}; what we add here is the demonstration that the scaling of $P_{\rm mod}$ with respect to $P$ extends over a whole sample of potentially precessing pulsars, consistent with a similar level of strain in each.  

Secondly, we argue the sharp magnetospheric switching and precession are not mutually exclusive.  Our argument is qualitative, but, we believe, persuasive.  In brief, suppose some pulsar is delicately balanced between two magnetospheric states.  This would be the case if the energy available to accelerate particles is close to some threshold for some avalanche-like process, say pair production.  Then, if the star is precessing, at some precessional phases the critical energy for the process is exceeded, favouring one magnetospheric state, while at other precessional phases the critical energy is not exceeded, favouring the other magnetospheric state.  That some pulsars should be so delicately balanced is plausible, given the random nulling and mode changing observed even in non-precessing pulsars.  The precession simply adds a bias to the probability of the magnetosphere being in one state or the other.  The corresponding changes in spin-down torque would then serve to amplify the effect of precession on the timing data: it would then be no surprise that switching behaviour and precession have been observed to go hand-in-hand. 

It is this precession hypothesis that we explore in this paper, whose structure  is as follows.  In section \ref{sect:observations} we give a summary of the relevant observations.  In section  \ref{sect:basic_model} we set out our model.  In section  \ref{sect:tests} we describe some consistency tests that we have applied to our model.  In section  \ref{sect:discussion} we discuss our results, and comment on some alternative explanations.  Finally, in section  \ref{sect:summary} we summarise our findings and give some concluding remarks.  Where relevant, we have assumed neutron star masses of $M = 1.4 M_\odot$, radii $R = 10^6$ cm, and moments of inertia $I = 10^{45}$ g cm$^2$.  We have made use of the ATNF pulsar database, as described in \citet{metal05}\footnote{\texttt{http://www.atnf.csiro.au/research/pulsar/psrcat}}.

\section{Summary of relevant observations} \label{sect:observations}

\citet{ketal06}  have reported on PSR B1931+24, which has the unique property of displaying large and correlated changes in spin-down rate and radio pulse strength: the pulsations are either `on' or `off', and the star seems to spin-down approximately 50\% faster when on than when off.  The switching between states occurs sharply--on a timescale of less than 10 seconds.  This has a natural interpretation as being caused by some  magnetospheric process, which is either on, contributing to the spin-down and providing the radio pulsations, or off,  contributing to neither spin-down nor radio pulsation \citep{ketal06}.

More recently, \citet{hlk10}  have presented the results of a study of timing noise in 366 pulsars, using data sets gathered at Jodrell Bank.  The long duration of the data sets ($\sim$ several decades) enabled them to identify pulsars whose timing residuals displayed interesting long-period ($\sim$ several year) harmonic structure.  This theme was taken up in more detail by \citet{letal10}.  The following interesting features emerged: (i) A subset of 17 pulsars displayed significant harmonic variation in their spin-down rate (see Fig 2 of \citet{letal10}).   (ii) In at least six of these pulsars, the variations in the spin-down rate were correlated with variations in pulse profile (see Fig 4 of \citet{letal10}).  On the basis of these two observations, one might be tempted to advance free precession as the mechanism responsible for the variations.  However, motivated by the peculiar properties of PSR B1931+24, \citet{letal10} analysed the pulsars further and found (iii) For at least two pulsars, the variations in pulse profile were smooth only when averaged over a sufficiently long timescale ($\sim 100$ days).  When studied on shorter timescales ($\sim 10$ minutes), the pulse profile showed signs of being in one of two states (see Fig 2 of \citet{letal10}), with the switch from one state to another occurring on a very short timescale; see Fig 5 of \citet{letal10}.  The longer-term smooth harmonic variation was then to be understood as being caused by the fraction of time the star spends in one state or the other smoothly varying on the long modulation timescale.  

This last point, \emph{viz} the sharp switching between pulse profiles, is clearly reminiscent of the behaviour of PSR B1931+24.  This in turn suggested that not only the pulse profile but also the spin-down torque itself in the set of $17$ pulsars was undergoing sudden sharp changes, caused by a similar sort of magnetospheric switching process as in B1931+24, although it was not possible to confirm this as torque variations can be detected only via variations in spin-down rate, and the spin-down could only be measured by averaging over timescales much longer than the switching timescale.  In any case,  the precessional hypothesis, with its smoothly varying modulation in pulse profile and spin parameters (see below) was rejected in favour of a magnetospheric switching hypothesis, which more naturally seems to accommodate sharp changes in spin-down torque and pulse profile \citep{letal10}. 

However, it remains the case that long term periodicities are seen in the spin-down of the 17 pulsars discussed by \citet{letal10}, and that some  have such clear timing variations that one can easily pick them out by eye, e.g. see the plots of spin-down rate for PSRs B1540-06, B1642-03, B1826-17 and B1828-11 in Figure 2 of \citet{letal10}.  There is no obvious mechanism to provide the clock for this within the magnetospheric switching hypothesis as formulated above.  Given this deficiency,  we will now reexamine the precession hypothesis as an explanation for the timing data of \citet{letal10}, paying particular attention to whether or not free precession is likely to be consistent with the sharp switching in pulse amplitude seen in some of the pulsars.

%%%%%%%%%%%%%%%%%%%%%%%%%%%%%%%%%%%%%%
\section{Basic model}  \label{sect:basic_model}
%%%%%%%%%%%%%%%%%%%%%%%%%%%%%%%%%%%%%%

Let us suppose that free precession is indeed the source of the harmonic structure seen in the timing data.  That free precession should produce variations in pulse arrival time is well known  \citep{rude70, gold70}.  Following \citet{gold70}, if a neutron star is (for the sake of simplicity) idealised as a biaxial body with principal moments of inertia $[I_1, I_1, I_3]$,  its angular velocity vector $\Omega_i$ evolves according to
\begin{eqnarray}
I_1 \dot \Omega_1 + \Omega_2 \Omega_3 (I_3-I_1) & = & T_1 ,\\
I_1 \dot \Omega_2 - \Omega_2 \Omega_3 (I_3-I_1) & = & T_2 ,\\
I_3 \dot \Omega_3  & = & T_3
\end{eqnarray}
where $T_i$ is the spin-down torque.  The motion of the body can then be understood using a method of successive approximations.  To leading order the motion is the well known free precession, found by setting $T_i = 0$ in the above.  Assuming that the body is close to spherical ($(I_3-I_1)/I_1 \ll 1$) and that the wobble angle $\theta$ (the constant angle between the body's $3$-axis and its fixed angular momentum vector) is small ($\theta \ll 1$) one obtains
\begin{eqnarray}
\label{eq:Omega_1}
\Omega_1 &=& \Omega \theta \cos \psi ,\\
\label{eq:Omega_2}
\Omega_2 &=& \Omega \theta \sin \psi ,\\
\label{eq:Omega_3}
\Omega_3 &=& \Omega ,
\end{eqnarray}
where $\Omega = (\Omega_i \Omega_i)^{1/2}$ is the magnitude of $\Omega_i$, $\psi$ is a phase that tracks the slow precession motion
\begin{equation}
\psi =  \Omega_{\rm fp} t ,
\end{equation}
$\Omega_{\rm fp}$ the slow precession frequency
\begin{equation}
\Omega_{\rm fp} =  \epsilon_{31}  \Omega ,
\end{equation}
$\epsilon_{31}$ the fractional asymmetry in the moment of inertia tensor
\begin{equation}
\epsilon_{31} = \frac{I_3-I_1}{I} \ll 1 ,
\end{equation}
and $I $ denotes the stellar moment of inertia, neglecting the small difference between different axes.

This torque-free precession produces variations in the timing and pulse profiles.  The apparent spin frequency is modulated by a fractional amount $\epsilon_{31}  \theta$  once per precession cycle $P_{\rm fp} = 2\pi/\Omega_{\rm fp}$, while the pulse shape will be modulated by the pulse emission cone wandering over a latitudinal angular interval $\pm \theta$ about its average location, relative to a fixed observer \citep{rude70}.

This torque-free motion can then be inserted into the $T_i \neq 0$ equations of motion given above to compute the motion to the next order of accuracy.  The variation in spin frequency is then
\begin{equation}
\label{eq:d2obdts}
\frac{d}{dt} (\Omega^2) = \frac{2}{I} \left[ T_i \Omega_i - \frac{I_3-I_1}{I} T_3 \Omega_3 \right] ,
\end{equation}
\citep{gold70}.  As was emphasised by \citet{cord93}, the torque, in addition to producing the well-known secular spin-down, introduces additional timing variations: the free precession modulates the spin rate and the latitudinal position of the magnetic axis.  Assuming that the spin-down torque is sensitive to such changes, this produces a variability in the torque, i.e. a time variation in the torque locked in phase with the precession, which in turn modulates the spin-down rate.  The precise form of the spin-down variation depends upon the functional form of $T_i$.  Schematically, we can neglect the small last term of equation (\ref{eq:d2obdts}) to give the (rather obvious) relation
\begin{equation}
\label{eq:dOd}
\delta \dot \Omega(t) \approx  \frac{\delta T[\phi(t)]}{I} ,
\end{equation}
where $\delta T[\phi(t)]$ denotes that part of the torque that is modulated by the precessional motion.
We have written $\delta T = \delta T[\phi(t)]$ to make it clear that the modulation is a function of precessional phase, which is itself a linear function of time.  Integrating with respect to time gives the corresponding variation in spin frequency
\begin{equation}
\label{eq:dO}
\delta \Omega(t) = \int^t  \frac{\delta T[\phi(\hat t)]}{I} d\hat{t}
\end{equation}
and integrating again gives the phase residuals:
\begin{equation}
\label{eq:dP}
\delta \Phi(t) = \int^t \delta \Omega(\hat{t}) \,  d\hat{t} .
\end{equation}

If the variation with torque is harmonic, i.e. if $\delta T$ is an oscilliatory function of time with frequency $\Omega_{\rm fp}$,  these integrals with respect to time simply introduce factors of the free precession period $\Omega_{\rm fp}$
\begin{equation}
\delta\dot\Omega \approx \frac{\delta T}{I} \Rightarrow \delta\Omega \approx \frac{\delta T}{I \Omega_{\rm fp}} \Rightarrow 
\delta\Phi \approx \frac{\delta T}{I \Omega_{\rm fp}^2} . 
\end{equation}
These equations can be rewritten in a  more illuminating form.  Defining a  spin-down timescale $\tau_{\rm sd} = \Omega/(2\dot\Omega)$, and writing in terms of the fractional variation in torque $\delta T/T$ and free precession period $P_{\rm fp} = 2\pi/\Omega_{\rm fp}$, we obtain
\begin{equation}
\label{eq:residuals}
\frac{\delta\dot\Omega}{\dot\Omega} \approx \frac{\delta T}{T} 
\Rightarrow 
\frac{\delta\Omega}{\Omega} \approx  \frac{1}{4\pi} \frac{\delta T}{T}  \frac{P_{\rm fp}}{\tau_{\rm sd}} 
\Rightarrow 
\delta\Phi \approx  \frac{1}{4\pi}  \frac{\delta T}{T}  \frac{P_{\rm fp}^2}{P \tau_{\rm sd}} .
\end{equation}

The torque itself is believed to be composed of two parts: a part from direct emission of electromagnetic waves from the time-varying magnetic multipole moments of the star, and a part caused by electromagnetic emission from charged particles being accelerated in the magnetosphere \citep{gj69, rs75}.  

The direct electromagnetic emission from the star's magnetic multipoles is probably dominated by the magnetic dipole, explicit forms of which have been worked out previously \citep{gold70, mela00}.    In any case, for each contributing multipole, the associated spin-down torque will be some function of the time derivatives of a `magnetic axis' $m_i$, fixed in the star, and the time derivatives themselves are generated by the rotation $\Omega_i$:
\begin{equation}
\frac{d^{n+1}m_i}{dt^{n+1}} = \epsilon_{ijk} \Omega_j \frac{d^{n}m_k}{dt^{n}} .
\end{equation}
The precessional motion will smoothly modulate these time derivatives (see equations (\ref{eq:Omega_1})--(\ref{eq:Omega_3})), so that this piece of the spin-down torque will be a smooth function of the precessional phase and therefore a smooth function of time.  We can then make a simple estimate $\delta T / T \sim \theta$, leading to 
\begin{equation}
\frac{\delta\dot\Omega}{\dot\Omega} \approx \theta
\Rightarrow 
\frac{\delta\Omega}{\Omega} \approx  \frac{1}{2\pi} \theta  \frac{P_{\rm fp}}{\tau_{\rm sd}} 
\Rightarrow 
\delta\Phi \approx  \frac{1}{2\pi} \theta  \frac{P_{\rm fp}^2}{P \tau_{\rm sd}} .
\end{equation}
It was equations of this form, adjusted to reflect the exact functional form of the magnetic dipole spin-down torque,  that were used by \citet{ja01} to extract wobble angles $\theta$ from the timing variations of precession candidates.   

Now let's consider the magnetospheric contribution to the spin-down torque.    Unfortunately, the magnetospheric torque is based on intrinsically more complicated physics than the direct electromagnetic emission, so explicit formulae for the torque as a function of stellar magnetisation and spin rate are not available.  Nevertheless, it is straightforward to argue that the magnetospheric torque should have a part locked in phase with the precession.  Ultimately, the magnetospheric torque requires the extraction of charged particles from the stellar surface by the strong electric fields $E_i$ generated by the rotation of the magnetic field $B_i$ embedded in the highly conducting star \citep{gj69, rs75}.  Assuming perfect conductivity at the stellar surface, the induced electric field is
\begin{equation}
{\bf E} = -\frac{1}{c} {\bf v} \times {\bf B} ,
\end{equation}
where ${\bf v} = {\bf \Omega} \times {\bf r}$ is the velocity of the stellar surface, i.e.
\begin{equation}
{\bf E} = -\frac{1}{c} ({\bf \Omega} \times {\bf r})    \times {\bf B} .
\end{equation}
Given that, to an excellent approximation, $B_i$ will be fixed with respect to the star, the dependence of $\Omega_i$ on precessional phase guarantees the precession-sensitive nature of this accelerating electric field.  This modulation in the accelerating field will inevitably produce modulation in the energies to which particles are accelerated in the magnetosphere, and therefore also to modulation in the spin-down torque, although the link between these depends upon the exact and poorly understood magnetospheric mechanism at work.  It follows that even in the case of magnetospheric torques, the spin-down rate will be modulated by the precession motion.  

As described in section \ref{sect:observations}, there is evidence for sharp switching behavior in the pulsar profile, which has been argued to correspond to sharp changes in the spin-down torque.  How can this be reconciled with free precession, where the spin variation of equations (\ref{eq:Omega_1}--\ref{eq:Omega_3}) above are smooth functions of time?  In the case of spin-down by direct electromagnetic emission, there is clearly no way of producing sharp changes in torque: any magnetic multipole tied to the star's precessional motion will have a smooth time variation (as viewed from the inertial frame) and so will produce a smooth variation in torque.    In the case of magnetospheric torque, the situation is less clear.    One might be tempted to argue that the observed switching  represents the precession pushing the available particle energy at some point in the magnetosphere over some critical threshold for some emission process to occur.  For those precessional phases where the particle energy exceeds the threshold, the process occurs, while for those precessional phases where the energy is too low, the process cannot occur.

However, this in itself will not lead to sharp switching, as presumably the \emph{area} of stellar surface over which the threshold is exceeded is itself a smoothly varying function of precession phase, going from zero up to some maximum and smoothly back to zero once per precession cycle.  What is needed for the sharp switching is that, once the energy threshold is exceeded at a single point, the process rises to a significant level, i.e. there is an `avalanche' effect.  Pair production is a clear candidate for providing this avalanche behaviour.   The energy threshold would be that for pair production itself, related to the non-zero rest mass of the electron/position pair, although there are other energy thresholds that might be relevant, such as the work function for particle extraction from the stellar surface.

That the magnetosphere might be so delicately balanced is made plausible by the well known phenomenon of mode changing and nulling \citep{lg06}.  These  are well established phenomenon, showing that even in non-precessing stars the magnetosphere can be delicately balanced between a few quasi-equilibrium states, with random changes between the states occurring.  The addition of precession will then presumably weight the magnetosphere toward one state or another, depending upon the precessional phase.  That the switching in precessing stars is only quasi-randomly related to the precession phase is then a natural consequence of the random nature of mode switches in non-precessing stars.  Our proposal is that the precession then causes the star to move between phases where it is more likely (as compared to average) to be in one magnetospheric state and  phases where it is less likely to be in the state.

In our precession model this corresponds to a component $\delta T$ of the torque being either on or off, with a probability tied to the precessional phase.  In this case we no longer have $\delta T \sim \theta T$;  the size of the torque variation is determined not by the precession angle $\theta$ but by the magnitude of the process being switched on/off in the magnetosphere.  The wobble angle only needs to be large enough to carry the star from below the threshold to over the threshold (and back) once per precession cycle.  It is still the case that equations (\ref{eq:dOd})--(\ref{eq:dP}) apply, but now $\delta T$ is some function only loosely tied to the precessional phase $\psi$.  That such a fluctuating torque can account for real timing data was demonstrated  by \citet{letal10} for PSR B1828-11; see their Figure S1.  Given that this is an additional source of harmonic variation, over and above the smooth variation, this will tend to make precession more noticeable in pulsars with switching magnetospheric states.  

To sum up, our model is as follows.   The basic mechanism at work in creating the periodicities seen in timing data is free precession.  This motion induces modulations in the spin down torque.  Part of this modulation is a smooth function of precessional phase.  However, part of the spin-down torque  is either on or off, the on state corresponding to the energy threshold of some avalanche-like process being exceeded at some point on the stellar surface or magnetosphere.

%%%%%%%%%%%%%%%%%%%%%%%%%%%%%%%%%%%%%%
\section{Comparison with observations}  \label{sect:tests}
%%%%%%%%%%%%%%%%%%%%%%%%%%%%%%%%%%%%%%

Having constructed a simple free precession hypothesis, we now attempt to evaluate it in the light of recent pulsar observations.  We will draw our data set from the pulsars whose timing features were studied in detail in \citet{letal10}.  \cite{letal10} reported on $18$ pulsars in total; $17$ taken from \citet{hlk10}, and also the intermittent pulsar PSR B1934+21, first reported in \cite{ketal06}.   For each pulsar we computed the quantity
\begin{equation}
\label{eq:N}
N = \frac{T_{\rm s}}{P_{\rm mod}} ,
\end{equation}
where $T_{\rm s}$ is the time-span of the observations and $P_{\rm mod}$ the reported modulation period.  The quantity $N$ is therfore the number of modulation cycles that fit into the total data span, and is (one) measure of how convincing the evidence is for harmonic structure in the timing data.  Given that the data have been fit to timing models that include the pulsar spin frequency $\nu$, frequency derivative $\dot \nu$, and (in some cases) second derivative $\ddot \nu$, but no higher derivatives, some low order polynomial terms will inevitably remain in noisy data, even in the absence of precession, making the interpretation of low-$N$ oscillations as real features dangerous.  We will therefore make use of  those pulsars from \citet{letal10} that have $N>3$; the exact cut-off of $3$ is somewhat arbitrary.  There are 15 such pulsars; we are not including pulsars  B2035+36 ($N=0.37$), J2043+2740 ($N=1.1$) or B0950+08 ($N=2.6$) in our sample.  We give, in Table \ref{table:pulsar_data} below, data for  these 15 pulsars, taken from \citet{ketal06}, \citet{hlk10} and \citet{letal10}.

\begin{table*}
 \begin{minipage}{115mm}
 \caption{Pulsar data used in our analysis.  The pulsar spin period $P$ and modulation period $P_{\rm mod}$  are taken from \citet{letal10}, while the observation timespans $T_{\rm s}$ are taken from \citet{hlk10},  with the exception of  PSR B1931+21 whose timespan  was taken from \citet{ketal06}.  The quantity $N$, defined in equation (\ref{eq:N}), is a measure of the significance of the reported modulations.  The quantity $T_{\rm s} \Delta F$ is a measure of the stability of the modulation period, as discussed in section \ref{sect:stability}; $\Delta F$ is the uncertainty in the modulation frequency, taken from \citet{letal10}.  }
 \label{table:pulsar_data}
 \begin{tabular}{@{}lllllll}
 Pulsar name & $P$ (seconds) & $P_{\rm mod}$ (years) & $P/P_{\rm mod}$ & $T_{\rm s}$ (years) & $N=T_{\rm s}/P_{\rm mod}$ & $T_{\rm s} \Delta F$ \\
B0740-28 & $0.167$ & $0.370$ & $1.425e-08$ & $20.60$ & $55.62$ & $4.12$ \\ 
B0919+06 & $0.431$ & $1.613$ & $8.450e-09$ & $27.00$ & $16.74$ & $1.08$ \\ 
B1540-06 & $0.709$ & $4.167$ & $5.386e-09$ & $19.70$ & $4.73$ & $0.39$ \\ 
B1642-03 & $0.388$ & $3.846$ & $3.190e-09$ & $35.70$ & $9.28$ & $2.50$ \\ 
B1714-34 & $0.656$ & $3.846$ & $5.399e-09$ & $15.20$ & $3.95$ & $0.61$ \\ 
B1818-04 & $0.598$ & $9.091$ & $2.082e-09$ & $35.10$ & $3.86$ & $0.35$ \\ 
B1822-09 & $0.769$ & $2.500$ & $9.737e-09$ & $19.70$ & $7.88$ & $1.38$ \\ 
B1826-17 & $0.307$ & $3.030$ & $3.207e-09$ & $18.20$ & $6.01$ & $0.36$ \\ 
B1828-11 & $0.405$ & $1.370$ & $9.357e-09$ & $18.70$ & $13.65$ & $0.37$ \\ 
B1839+09 & $0.381$ & $1.000$ & $1.207e-08$ & $25.30$ & $25.30$ & $3.79$ \\ 
B1903+07 & $0.648$ & $2.778$ & $7.383e-09$ & $19.60$ & $7.06$ & $2.55$ \\ 
B1907+00 & $1.017$ & $6.667$ & $4.829e-09$ & $23.60$ & $3.54$ & $0.47$ \\
B1929+20 & $0.268$ & $1.695$ & $5.008e-09$ & $17.20$ & $10.15$ & $0.34$ \\
B1931+24 & $0.814$ & $0.076$ & $3.373e-07$ & $1.67$ & $21.88$ & $1.17$ \\ 
B2148+63 & $0.380$ & $3.030$ & $3.969e-09$ & $19.80$ & $6.53$ & $1.39$
\end{tabular}
\end{minipage}
\end{table*}

We will use these pulsars to examine our precessional hypothesis.

%%%%%%%%%%%%%%%%%%
\subsection{Explaining the modulation period}  \label{sect:etmp}

In the precession interpretation, we identify $P_{\rm mod}$ with $P_{\rm fp}$, the free precession period of a biaxial star, 
as described in section \ref{sect:basic_model}, modulo allowance for the fact that real neutrons star are not simple rigid bodies.  When allowance is made for the fact that a real star consists of an elastic crust containing a fluid core, the ratio of spin to free precession periods becomes
\begin{equation}
\frac{P}{P_{\rm fp}} = \frac{I_3-I_1}{I_{\rm C}}
\end{equation}
where $I_3-I_1$ refers to that part of the asymmetry in the moment of inertia tensor, generated by strains, that follows the precessional motion, while $I_{\rm C}$ refers to the crust plus all other parts of the star that are coupled to it on timescales shorter than the spin period (see \citet{ja01} and references therein).  

The asymmetry $I_3-I_1$ we will continue to parameterise as a dimensionless number 
\begin{equation}
\epsilon_{31} = \frac{I_3-I_1}{I_\star}
\end{equation}
where $I_\star$ denotes the total stellar moment of inertia.  The size of $\epsilon_{31}$  will depend upon the strain mechanism at work, i.e. whether it is elastic or magnetic, as will be discussed below.

The size of $I_{\rm C}$ requires some thought, as its value depends on the coupling between the crust and the rest of the star.  If the entire star is threaded by a magnetic field,  then the interior charged plasma (electrons and protons) will be tightly coupled to the crust \citep{eass79}, with a coupling timescale much shorter than any of the timescales of interest here.  The crust and interior plasma can then be thought of as a single component, so that $I_{\rm C}$ at the very least consists of the moment of inertia of the crust, plus a small extra piece contributed by the interior plasma.  

It is then necessary to consider the interaction between the crust/plasma and the core neutron fluid.  In the case that
the core neutrons are normal (i.e. not superfluid) then they will be strongly coupled to the charged plasma in a frictional-like way \citep{bpp69}.  The case of a two-component star, with its components coupled by a frictional torque proportional to the vectorial angular velocity difference, was considered by \citet{bg55}, who showed than if this  frictional coupling acts on a timescale of order the spin period $P$ or less, the whole star, curst plus fluid core, will precess as one.  Conversely, if the frictional coupling acts on timescales much greater than $P$, then the fluid core does not participate in the precession.  It follows that in the case that the core neutrons are normal,   the whole star, core plus crust, precesses as one, giving $I_{\rm C} \sim I_\star$.  

On the other hand, if the interior neutrons are superfluid, the interaction between crust and core is mediated by a process known as mutual friction.  This can be provided by the charged plasma scattering off the neutron rotational vortices \citep{as88}.  The precession of a two-component star, with the components coupled by a mutual friction torque, was considered in detail by \citet{swc99}.  In the limit of sufficiently weak mutual friction, a long period precessional mode still exists, with the neutron fluid decoupled from the precession, so that  $I_{\rm C}$ would then  be of order the crustal moment of inertia, i.e. $I_{\rm C}/I_\star \sim 10^{-2}$ \citep{swc99, gaj08, gaj09}.  
 
However, it has been argued that there will be an even stronger interaction between the crust and fluid core if the interior protons are superconducting: then the magnetic flux tubes that make up the superconducting field will interact strongly with the neutron vortices.  Link has argued that this interaction is so strong that  there will be no long-lived long period free precession, forcing one to conclude that either the pulsar observations do not correspond to precession, or else the neutron vortices do not coexist with the magnetic flux tubes \citep{link03, link06}.  In the latter case, one might conclude that the interior neutrons are normal, returning us to the $I_{\rm C} / I_\star \sim 1$ case described above.  Alternatively, one might conclude that the neutron vortices are spatially separate from the precessing crust and magnetic field, decoupling then from the precession, implying $I_{\rm C}/I_\star \sim 10^{-2}$.   Note that the recent cooling observations of Cas A \citep{hh10} suggest that both neutron superfluidity and proton superconductivity are present in all but the youngest of neutron stars \citep{petal11, setal11}, favouring the second of these suggestions.

Given the uncertainty in the correct value of $I_{\rm C}/I_\star$, we will explicitly introduce the ratio $I_{\rm C}/I_\star$ in our calculations:
\begin{equation}
\label{eq:P_over_P_fp}
\frac{P}{P_{\rm fp}} = \frac{I_3-I_1}{I_{\rm C}}  = \frac{I_3-I_1}{I_\star} \frac{I_\star}{I_{\rm C}} = \epsilon_{31}
\left(\frac{I_{\rm C}}{I_\star}\right)^{-1} ,
\end{equation}
and consider both the cases $I_{\rm C}/I_\star \approx 10^{-2}$ and $I_{\rm C}/I_\star \approx 1$ in our discussions.

It remains to consider the value of the ellipticity parameter $\epsilon_{31}$.  For an unmagnetised fluid star $\epsilon_{31}$ is zero; non-zero $\epsilon_{31}$ values are generated only by strains, which may be either elastic or magnetic in origin.  We will now consider the different possibilities.

Let's first consider elastic strains, which will be supported by the solid crust.  For a star whose crust is relaxed when it has a shape parameterised by $\epsilon_0$, the `zero strain' or `reference' oblateness,  we expect
\begin{equation}
\epsilon_{31} \approx b \epsilon_0 ,
\end{equation}
where $b$ is a measure of the rigidity of the crust \citep{ps72, ja01}. In \cite{cul03} it is estimated that $b \sim 10^{-7}$ for typical neutron stars.  

For a star with a completely relaxed crust, its reference shape will be the shape of the equivalent rotating fluid, which has an oblateness of order the kinetic energy divided by the gravitational binding energy:
\begin{equation}
\epsilon_\Omega \approx \frac{MR^2\Omega^2}{GM^2/R} \approx  2.1 \times 10^{-7} \left(\frac{1 \, \rm s}{P}\right)^2 .
\end{equation}
It follows that for a star whose crust is perfectly relaxed at its current rotation rate we can set $\epsilon_0 = \epsilon_\Omega$ to give
\begin{equation}
\epsilon_{31} \approx b\epsilon_\Omega \approx 2.1 \times 10^{-14} \left(\frac{b}{10^{-7}}\right)
\left(\frac{1 \, \rm s}{P}\right)^2 ,
\end{equation}
which translates into a modulation period
\begin{equation}
\label{eq:P_mod_relaxed}
\frac{P_{\rm mod}}{1 \, \rm year}  \approx  1.5 \times 10^6  \left(\frac{10^{-7}}{b}\right)  \left(\frac{P}{1 \, \rm s}\right)^3 
\left(\frac{I_{\rm C}}{I_\star}\right) .
\end{equation}

In reality  we would expect a spinning-down star to have a larger value of $\epsilon_{31}$ than this, as the crust will be strained, retaining a memory of an earlier faster rotation rate. The exact picture depends rather subtly on the spin period at the moment when the crust first solidifies, and the breaking strain.  To understand why, it is useful to consider the following evolutionary picture.  Suppose the crust first solidifies when the star is  spinning fast.  The crust is in fact believed to form very shortly after birth, so we will label the ellipticity of this relaxed star as  $\epsilon_0 = \epsilon_{\Omega, \rm birth}$, with the understanding that, technically, `birth' refers to the formation of the crust, not the slightly earlier birth of the star.  As the star then spins down, a strain builds up; if the strain reaches the breaking value $u_{\rm break}$, a starquake occurs, resetting $\epsilon_0$ to a new value appropriate to the current rotation rate.  This process can occur many times, such that $\epsilon_0 \approx \epsilon_\Omega$ throughout.  However, when the star has spun down to the point that $\epsilon_0 \approx u_{\rm break}$, further spin-down will not generate sufficient strain to cause further starquakes, so that the strain is then `frozen in' at the level of $u_{\rm break}$.  Algebraically we have
\begin{eqnarray}
\label{eq:epsilon_fast}
\epsilon_{31} &\approx& b\epsilon_\Omega {\,\, \rm for \,\,} P < P_{\rm crit} ,\\
\label{eq:epsilon_slow}
\epsilon_{31} &\approx& b u_{\rm break}  {\,\, \rm for \,\,} P > P_{\rm crit} ,
\end{eqnarray}
where the critical dividing spin period $P_{\rm crit}$  is given by $\epsilon_\Omega (P_{\rm crit}) = u_{\rm break}$ so that
\begin{equation}
\label{eq:P_crit}
P_{\rm crit} \approx  1.5 \times 10^{-3} {\, \, \rm s \, \,} \left(\frac{10^{-1}}{u_{\rm break}}\right)^{1/2} ,
\end{equation}
where our parameterisation of $u_{\rm break}$ is motivated by the very high breaking strains computed recently by \citet{hk09}.  

However, if the breaking strain really is as high as this, the critical period $P_{\rm crit}$ of equation (\ref{eq:P_crit}) is probably shorter than the actual birth spin period of pulsars, invalidating the above evolutionary scenario: spin-down would not induce sufficient strain to break the crust and so  starquakes would not occur as the star spins down.  Instead, the zero strain oblateness remains fixed at its birth value, $\epsilon_0 = \epsilon_{\Omega, \rm birth}$, for the entire stellar lifetime.  In this case
\begin{equation}
\label{eq:epsilon_slow_birth}
\epsilon_{31} = b \epsilon_{\Omega, \rm birth}
\end{equation}
Once such a star has spun-down by a factor of a few, the level of strain would reach $u \approx \epsilon_{\Omega, \, \rm birth}$, below the breaking strain  $u_{\rm break}$.

Taking all this into account, we can parameterise the ellipticity:
\begin{eqnarray}
\label{eq:epsilon_fast_param}
\epsilon_{31} &\approx& 2.1 \times 10^{-14}  \left(\frac{b}{10^{-7}}\right) \left(\frac{1 \, \rm s}{P}\right)^2     {\,\, \rm for \,\,} P < P_{\rm crit} ,\\
\label{eq:epsilon_slow_param}
\epsilon_{31} &\approx& 1.0 \times 10^{-8}    \left(\frac{b}{10^{-7}}\right)   \left(\frac{u}{10^{-1}}\right)     {\,\, \rm for \,\,} P > P_{\rm crit} .
\end{eqnarray}
with corresponding  modulation periods given by
\begin{equation}
\label{eq:P_mod_fast}
\frac{P_{\rm mod}}{1 \, \rm year}  \approx  1.5 \times 10^6  \left(\frac{10^{-7}}{b}\right)  
\left(\frac{P}{1 \, \rm s}\right)^3  \left(\frac{I_{\rm C}}{I_\star}\right) {\,\, \rm for \,\,} P < P_{\rm crit} ,
\end{equation}
\begin{equation}
\label{eq:P_mod_slow}
\frac{P_{\rm mod}}{1 \, \rm year}  \approx   3.2 \left(\frac{10^{-7}}{b}\right) \left(\frac{0.1}{u}\right) \left(\frac{P}{1 \, \rm s}\right) \left(\frac{I_{\rm C}}{I_\star}\right)
{\,\, \rm for \,\,} P > P_{\rm crit} .
\end{equation}
with the strain $u$ in the $P > P_{\rm crit}$ case equal to
\begin{eqnarray}
u &\approx& u_{\rm break}   {\,\, \rm for \,\,} P_{\rm birth} < P_{\rm crit} ,\\
u &\approx& \epsilon_{\Omega, \, \rm birth}   {\,\, \rm for \,\,} P_{\rm birth} > P_{\rm crit} .
\end{eqnarray}

It follows that in the case of rapidly spinning star whose crusts can be approximated as relaxed we have $P_{\rm mod}\propto P^3$.  For slow spinning stars with a fixed level of strain, $P_{\rm mod} \propto P$, to the extent that all stars have identical breaking strains or identical birth spin periods.

In the case of magnetic field  deformation the situation is simpler.  In the case of a magnetic field of typical strength $B$ threading non-superconducting  stellar fluid, the expected ellipticity is of order the magnetic binding energy to the gravitational binding energy:
\begin{equation}
\epsilon_{31} \approx \frac{B^2 R^3}{GM^2/R} \approx 1.9 \times 10^{-12} \left(\frac{B}{10^{12} \, \rm G}\right)^2 ,
\end{equation}
which leads to
\begin{equation}
\label{eq:P_mod_normal}
\frac{P_{\rm mod}}{1 \, \rm year} \approx 1.6 \times 10^4 \left(\frac{10^{-15} \rm s \, s^{-1}}{\dot P}\right) 
\left(\frac{I_{\rm C}}{I_\star}\right) ,
\end{equation}
where we have used $B = 3.2 \times 10^{19} (P \dot P)^{1/2}$ (gaussian cgs units; see \citet{lg06}) and have parameterised using values of $B$ and spin-down rate $\dot P$ typical for the set of pulsars under consideration.

In the case of superconductivity these results are modified by the replacement of one factor of $B$ with $H_{\rm c}$, the `critical field', believed to approximately equal to $10^{15}$ G, independent of the star's individual field strength $B$ \citep{bpp69}.  The above results are then amended to
\begin{equation}
\epsilon_{31} \approx \frac{B H_{\rm c} R^3}{GM^2/R} \approx 1.9 \times 10^{-9} \left(\frac{B}{10^{12} \, \rm G}\right)
\left(\frac{H_{\rm c}}{10^{15} \, \rm G}\right) .
\end{equation}
 This leads to
\begin{equation}
\frac{P_{\rm mod}}{1 \, \rm year}
\approx 16  \left(\frac{P}{1 \, \rm s}\right)^{1/2} \left(\frac{10^{-15} {\rm s \, \rm s^{-1}}}{\dot P}\right)^{1/2} 
\left(\frac{10^{15} \, \rm G}{H_{\rm c}}\right) \left(\frac{I_{\rm C}}{I_\star}\right) ,
\end{equation}
or, in terms of spin-down age, $\tau_{\rm sd} = P/(2\dot P)$
\begin{equation}
\label{eq:P_mod_supercon}
\frac{P_{\rm mod}}{1 \, \rm year}  \approx 4.1 \times \left(\frac{\tau_{\rm sd}}{10^6 \, \rm years}\right)^{1/2}  
\left(\frac{10^{15} \, \rm G}{H_{\rm c}}\right) \left(\frac{I_{\rm C}}{I_\star}\right) .
\end{equation}
(As mentioned previously, given that we are attempting to model long period free precession, the existence of superconductivity is problematic, as a strong interaction between the magnetic flux tubes (not included in our model) that are believed to make up the magnetic field in a superconducting star with a coexisting neutron superfluid vortex array would tend to increase the precession frequency to very high values, or else damp the precession very rapidly \citep{ja01, link03, link06}.  A possible resolution would be for the neutrons to be normal.)

We therefore see that within the precessional hypothesis there are four different $P_{\rm mod}(P, \dot P)$ relationships, depending upon the mechanism at work: precession of an elastically relaxed star (equation (\ref{eq:P_mod_fast})), precession of an elastically strained star  (equation (\ref{eq:P_mod_slow})), precession of a magnetised but non-superconducting star (equation (\ref{eq:P_mod_normal})) or precession of a magnetised superconducting star (equation (\ref{eq:P_mod_supercon})).  
These are not all mutually exclusive as both elastic and magnetic strains can contribute to the overall strain, while the two quoted elastic results represents the expected extremes of crustal strain, if one sets the strain to its breaking value $u_{\rm break}$.

Clearly, the relaxed crust and normal magnetic field scenarios can be rejected immediately, as the predicted modulations periods of equations (\ref{eq:P_mod_fast}) and (\ref{eq:P_mod_normal}) are far in excess of these observed, regardless of where the ratio $I_{\rm C}/I_\star$ lies in the interval $10^{-2} \rightarrow 1$.  In contrast, in the cases of free precession of a star with a strained crust or a superconducting star, the periodicities of equations (\ref{eq:P_mod_slow}) and (\ref{eq:P_mod_supercon}) are of the right order of magnitude to explain the data, although in the latter case we would need $I_{\rm C}/I_\star \sim 1$.  To gain further insight we can make use of the different scaling of $P_{\rm mod}$ with $P$ and $\dot P$, and exploit the fact that  \cite{letal10} have collected data on multiple pulsars.

\begin{figure} 
\caption{\label{fig:P_mod_v_P} Plot of modulation period $P_{\rm mod}$ in years verses spin period $P$ in seconds for 15 pulsars from Table 1 of \citet{letal10}.  Error bars taken from Table 1 of \citet{letal10}.  The solid line is $P/P_{\rm mod} = 5 \times 10^{-9}$, and is not a fit to the data; a linear relation of this form is expected for the precession of stars with identical elastic strains generated by spin-down from an earlier faster rotation rate (see equation (\ref{eq:P_mod_slow}).} 
\includegraphics[angle=-90, width=90mm]{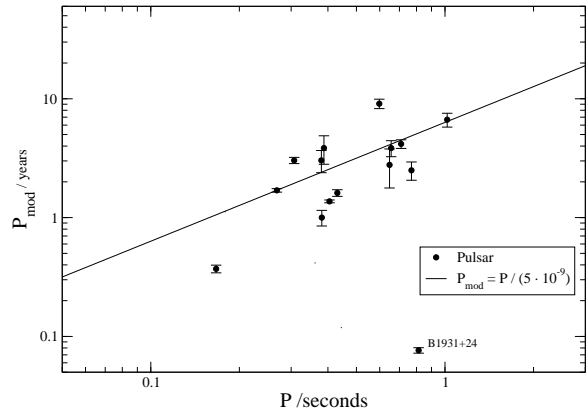} 
\end{figure}

In Fig \ref{fig:P_mod_v_P}  we plot the modulation period $P_{\rm mod}$ verses spin period $P$ for the pulsars given in Table \ref{table:pulsar_data}.  If the free precession of a strained star is at work, the data points should form a line of unit gradient;  to guide the eye we have  plotted a line corresponding to $P/P_{\rm mod} = 5 \times 10^{-9}$ (see equation (\ref{eq:P_mod_slow})).   Note the errors bars, taken from \citet{letal10}, measure the width of the power spectra peaks in the spin-down rate.
All but two pulsars lie within a factor of $2$ of the line  $P/P_{\rm mod} = 5 \times 10^{-9}$.  The exceptions are pulsar B0740-28 which lies a factor of  $3$ below the line, and the intermittent pulsar B1931+24, which lies a much larger factor of about $70$ below the line.

\begin{figure} 
\caption{\label{fig:P_mod_v_age} Plot of modulation period $P_{\rm mod}$ in years verses pulsar age $\tau = P/(2\dot P)$ in years for 15 pulsars from Table 1 of \citet{letal10}.  Error bars taken from Table 1 of \citet{letal10}.  The solid line is the expected scaling of $P_{\rm mod}$ for presession of superconducting stars deformed by magentic strains,  and is not a fit to the data.  It was obtained by setting $H_{\rm c}= 2 \times 10^{15}$ G in equation (\ref{eq:P_mod_supercon}).} 
\includegraphics[angle=-90, width=90mm]{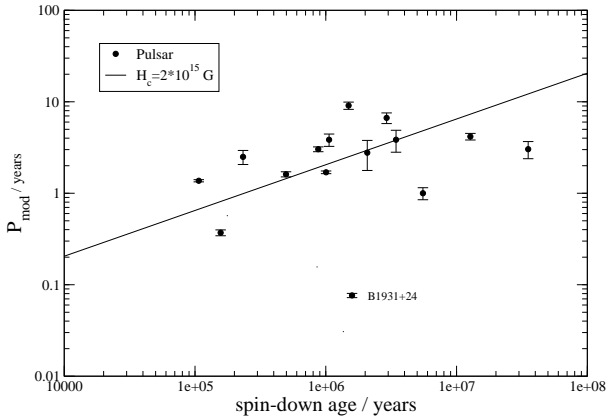} 
\end{figure}

In Fig \ref{fig:P_mod_v_age}  we plot the modulation period $P_{\rm mod}$ verses the pulsar age $\tau_{\rm sd} = P/(2\dot P)$.   If free precession of a star deformed by superconducting magnetic field is at work, the data should form a line of gradient $1/2$; to guide the eye, we have plotted such a line, corresponding to $H_{\rm c} = 2 \times 10^{15}$ G (see equation (\ref{eq:P_mod_supercon})). 
There is slightly more scatter about the indicated line in this case, with four pulsars lying a factor of $3$ or more off the line.  PSR B1931+24 is again a clear outlier, lying a factor of $30$ below the line.  Also, the $H_{\rm c} = 2 \times 10^{15}$ G  line appears to be somewhat steeper than the data.

To make a more quantitative statement as to the actual scaling of $P_{\rm mod}$ with $P$ and $\dot P$ we have carried out a least squares fit to the data, excluding  PSR B1931+24 due to its clear outlier status.   The scatter of the points about a straight line is likely to be dominated not be the measurement errors in $P_{\rm mod}$ but by intrinsic differences between the stars themselves, e.g. differences in mass (which affects crust thickness and hence $b$), differences in fracture history, or differences in internal magnetic field arrangement.  We can see no easy way of estimating the sorts of scatter these effects are likely to produce in the modulation period.   We will therefore not make use of any error estimates in calculating the best fit lines. Assuming a relation of the form $P_{\rm mod} = a P^b \dot P^c$, and fitting for $a, b, c$ leads to
\begin{equation}
\label{eq:best_fit}
\frac{P_{\rm mod}}{1 \, \rm year}
\approx \frac{(P/ 1 {\rm s})^{1.27\pm0.31} (\dot P / 10^{-15} {\rm s \, s}^{-1})^{-0.14\pm0.094}}{3.79 \times 10^{-9} (1\pm0.14)} .
\end{equation}
This scaling is close to that of equation (\ref{eq:P_mod_slow}), i.e. close to that expected for the free precession of a star deformed by elastic strains, with the level of strain approximately equal in all stars. Unfortunately our lack of knowledge of the intrinsic scatter in $P_{\rm mod}$ as a function of $P$ and $\dot P$ prevents us from making any statement as to the statistical significance of equation (\ref{eq:best_fit}).  All we can say is that of the four different deformation mechanisms proposed, the strained crust one is the most plausible.  Note that if we had assumed from the outset that $P_{\rm mod}$ was only a function of $P$,  and therefore carried out a fit of the form $P_{\rm mod} = a P^b$ the result would have been rather similar:
\begin{equation}
P_{\rm mod} \approx \frac{P^{1.26 \pm 0.33}}{4.8 \times 10^{-9} (1\pm0.13)} .
\end{equation}

Having seen that $P /P_{\rm mod} \approx 5\times 10^{-9}$ is a reasonable fit to that data, we can then use equation (\ref{eq:P_mod_slow}) to estimate the corresponding strain; obviously the actual braking strain must be at least as large as this:
\begin{equation}
u_{\rm break} \gtrsim u \approx 5.0 \times 10^{-2} \left(\frac{P/P_{\rm mod}}{5 \times 10^{-9}}\right) \left(\frac{10^{-7}}{b}\right)
\left(\frac{I_{\rm C}}{I_\star}\right)
\end{equation}
This lower bound on the breaking strain can then be translated, using equation (\ref{eq:P_crit}), into an upper bound on the critical spin period $P_{\rm crit}$; the birth spin period must equal to or less than this:
\begin{equation}
P_{\rm birth} \lesssim P_{\rm crit} \lesssim 2.1 {\, \rm ms\,} \left(\frac{b}{10^{-7}}\right)^{1/2} \left(\frac{I_{\rm C}}{I_\star}\right)^{-1/2}
\left(\frac{5 \times 10^{-9}}{P/P_{\rm mod}}\right)^{1/2}
\end{equation}

If the whole of the fluid interior participates in the precession, so that $I_{\rm C}/I_\star \approx 1$, we can therefore say $u_{\rm break} \gtrsim u \approx 5.0 \times 10^{-2}$ and $P_{\rm birth} \lesssim 21$ ms, or $f_{\rm birth} \gtrsim 490$ Hz.  This high level of strain is in fact consistent with the high breaking strains suggested by the work of \citet{hk09}.  However, this birth spin period is rather short: simple extrapolation of  the Crab pulsar back to its historically known birth date suggests a much longer birth period.  If instead only the crust participates in the precession, so that $I_{\rm C}/I_\star \approx 10^{-2}$, we obtain $u_{\rm break} \gtrsim u \approx 5.0 \times 10^{-4}$ and $P_{\rm birth} \lesssim 0.021$ s, or $f_{\rm birth} \gtrsim 49$ Hz, a more plausible bound on the birth spin rate.

In either case, the roughly uniform level of strain for all the precession candidates is interesting; the data points in Figure \ref{fig:P_mod_v_P} are scattered by a factor  $\sim 2$ about the best-fit line.  One possible explanation for the uniformity is simply that this reflects the actual breaking strain of neutron star crusts.  The larger value quoted above, $u=5 \times 10^{-2}$, is close to the $0.1$ value that has been obtained in the molecular dynamics simulations of \cite{hk09}.  As discussed in \citet{ch10}, when finite-temperature effects are taken into account, the breaking strain becomes a decreasing function of time.
Extrapulation to the long timescale relevant to precessing stars is dangerous, but a value of $u_{\rm break} \approx 5 \times 10^{-4}$ seems sensible.  The scatter by a factor of about $2$ about the best fit line would then correspond to minor differences in crust thickness and `geological' history between the different stars.  Alternatively, the uniformity in strain   may reflect similar spin periods at birth, with crust-only precession implying a birth spin of about $50$ Hz and whole-body precession implying a higher birth spin frequency closer to $500$ Hz. The scaling of $\epsilon_{\Omega, \, \rm birth}$ with the square of the birth spin frequency would require a scatter of no more than $\sim \sqrt 2$ about the average birth frequency, a tight clustering indeed.  Given this, it seems more plausible that the observed strain values really are giving us information about a universal neutron star crustal breaking strain, but a definitive statement cannot be made.

If we were to interpret the behaviour of PSR B1931+24 as free precession, despite its outlier status, we obtain the following.  For 
$I_{\rm C}/I_\star \approx 1$ we obtain the non-sensical strain $u=3.3$.  For $I_{\rm C}/I_\star \approx 10^{-2}$ we obtain $u=0.033$, which corresponds to a birth period of less than $2.6$ ms, or a birth frequency greater than $390$ Hz.  So, in this case, sensible results can only be obtained for crust-only precession.  If PSR B1931+24 is indeed undergoing free precession, we can provide no explanation as to why is has such an anomalously short precession period.

To sum up,  the relation between spin and modulation periods is of roughly the form to be expected for free precession in the regime where the crust retains a significant level of strain, built up from spin-down from a faster rotation rate, but PSR B1931+24 seems to have too short a modulation period to be described by the same model parameters as the rest of the data.

%%%%%%%%%%%%%%%%%%
\subsection{Stability of the modulation period}  \label{sect:stability}

Free precession should be a smooth regular motion, with a period set by the stellar structure, as described above.   A neutron star has a very high moment of inertia, so one expects the underlying clock to be extremely accurate.  Departures from  perfect periodicity in the modulations, if large enough, would make a precessional interpretation problematic.  There are, however, two complicating factors that need to be taken into account.  Firstly, the data are of finite duration, so some uncertainty in the extracted frequency is inevitable.  Secondly, the long-term periodic modulation is presumably not the only source of timing noise in the data; if sufficiently large, the other sources of timing noise can make extraction of the periodic modulation difficult, and possibly make it appear less regular than is actually the case.  Indeed, one such source of noise will be the stochastic nature of the magentospheric contribution to the spin-down torque.  It follows that only by averaging over sufficiently many modulation periods would one expect to recover the underlying high-precision clock.  This last point is in close analogy with the problem of measuring pulsar spin periods, as most pulsars display large pulse-to-pulse shape profile variations, and only by averaging over many can an average profile be constructed, then allowing the extraction of the highly stable spin period.

To investigate the stability of the modulation,  we can examine the power spectra of the spin-down rate as given in Figure S2 of \citet{letal10}.   The spectra for most stars consist of multiple-peaks, and all peaks are of finite width in frequency space.  This in itself is not necessarily a sign of wandering in the modulation frequency: observation of a periodic system over a finite time span will generically produce such power spectra.   Even in the case where the process under consideration is perfectly periodic, the width of these peaks is $\sim T_{\rm s}^{-1}$, where $T_{\rm s}$ is the observation span.  This is to be compared with the actual peak widths, $\Delta F$, as given in Table 1 of \cite{letal10}.  A convenient quantity to consider is then the ratio of these two frequency widths, $T_{\rm s} \Delta F$; a value of $T_{\rm s} \Delta F$ significantly in excess of unity would imply that the finite duration of the data span is \emph{not} the main factor determining the width of the frequency peaks.  We give this quantity in Table \ref{table:pulsar_data}.

As is clear from the table, the width of the vast majority of peaks is about that one would expect if the finite duration of the observations is the dominant source of frequency uncertainty.  The only pulsars with $T_{\rm s} \Delta F > 3$ are  B0740-28, and B1839+09, whose peak widths are about four times $T_{\rm s}^{-1}$.  On the basis of these power spectra it seems reasonable to say that in all or most cases there is no clear signature of wandering in the modulation period, and the data seem to be consistent with a highly stable underlying clock.  That this is true for PSR B1931+24  was implicit in the analysis of \cite{retal08}, who attempted to explain the behaviour of this system by invoking a companion object, a planet or low mass star.  Note, however, that Shabanova has reported some variation in pulse phase for PSRs B0919+06 and B1642-03 \citep{shab10, shab09}.  For now we will simply note that these pulsars do not have high  values of $T_{\rm s} \Delta F$ ($1.1$ and $2.5$, respectively) so by this measure at least are stable.  We will comment on this further in section \ref{sect:sg} when we discuss slow glitches.

%%%%%%%%%%%%%%%%%%
\subsection{Excitation of harmonic timing noise}  \label{sect:aotn}

What  makes the subset of pulsars given in Table \ref{table:pulsar_data} special, i.e. why is it these pulsars, and not others, display interesting harmonic structure in their spin-down rates? 
\begin{figure} 
\caption{\label{fig:P-P_dot_glitch} $P$--$\dot P$ diagram for the pulsar population, with the subset from Table \ref{table:pulsar_data} identified.  For comparison, we have also indicated those pulsars that have been observed to glitch.  Data taken from \citet{letal10} and the ATNF database.} 
\includegraphics[angle=-90, width=90mm]{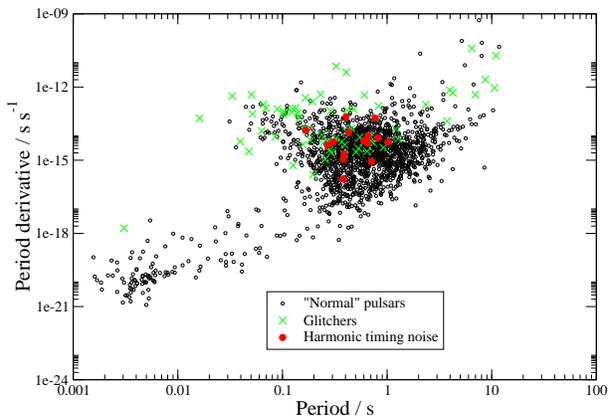} 
\end{figure}
To address this issue we plot a $P$--$\dot P$ diagram in Figure \ref{fig:P-P_dot_glitch}, where we plot `normal' pulsars, and also the pulsars with harmonic timing residuals.  For purposes of comparison, we have also  identified those pulsars that have been observed to glitch.

Indeed, one possible explanation for why only a small subset of pulsars display harmonic timing residuals might be related to glitches: perhaps the precession was excited by a recent glitch.  In PSR B1828-11, where the periodicity stands out cleanly, we can see that there has been no significant decay in the long-term modulation over the observation span.  This implies that the stars in our subset might have been set into precession decades ago, quite possibly before the first pulsar observations were taken.  However, this simple glitch model is clearly not correct: as it apparent from the $P$--$\dot P$ diagram, the vast majority of pulsars  that have been observed to glitch do not display harmonic timing structure.  Clearly, it is not the case that a glitch automatically leads to harmonic timing structure.  

However, an alternative explanation suggests itself, which makes use of the switching behaviour that is present in at least some of the precession candidates.  As explained above, precession with wobble angle $\theta$ will, in the case of a smoothly-varying torque function, produce torque variations of fractional order $\theta$.  However, for those special pulsars which are delicately balanced between the magnetospheric states, an additional fractional variation $\delta T/ T$ is present, where $\delta T$ is the difference in spin-down torques between the two magnetospheric states.  If this difference is sufficiently large, the free precession will be much more apparent in the timing data of these stars than in those that are not delicately balanced between magnetospheric states: \emph{the switching serves to magnify the torque variations}.  This suggests the following explanation for the apparent rarity of observed precession: those 15 or so candidates are special by virtue of being finely balanced between different states.  The observed switching is then not an argument against free precession, but in fact the mechanism that magnifies its effects to make it apparent in the radio timing data.

Of course, this model requires some non-zero level of precession to provide the clock to regulate the switching mechanism.  Unfortunately, we see no way of estimating how small a level suffices to make the switching mechanism work.  Certainly, if the balance between the two magnetospheric states is sufficiently delicate, the required precessional amplitude may be sufficiently small so the low level of undetected precession that presumably lies buried in regular pulsar timing noise might suffice.   That timing noise should indeed contain some precessional component is natural: whatever mechanism is responsible for producing timing noise, it is likely to excite some small degree of free precession \citep{cord93}.  This would only fail to be the case if the mechanism produced only perfectly axisymmetric variations in moment of inertia or torque fluctuations which only had a component along the rotation axis.  Alternatively, one can invoke a recent glitch to provide the necessary precession.  The lack of visible precession in the known glitching pulsars is then understood by none of them being sufficiently delicately balanced between different magnetospheric states.

\begin{figure} 
\caption{\label{fig:s2-age} Plot of pulsar rms timing noise $\sigma_2$ verses pulsar age $P/(2\dot P)$.  Data taken from \citet{hlk10}, where this quantity is labelled $\sigma_2$.  The $15$ pulsars of Table \ref{table:pulsar_data} with interesting harmonic structure in their timing properties are identified.} 
\includegraphics[angle=-90, width=90mm]{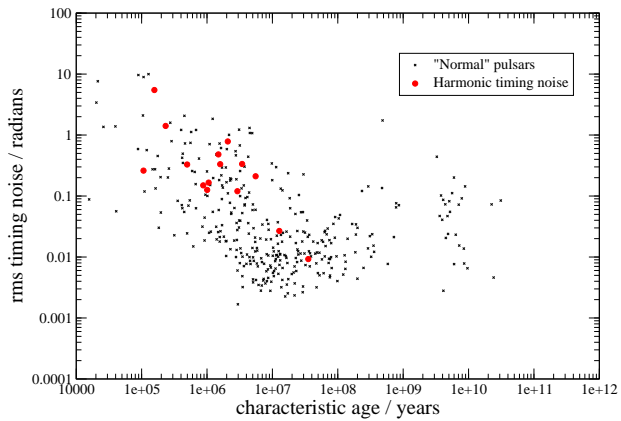} 
\end{figure}

It would then follow that the amount of timing noise should be greater in the precessing subset than for the other pulsars (or at least in those precessing stars where sharp changes in spin-down occur), as this subset has the conventional non-precessing timing noise contributions present in all pulsars, plus the small smooth precession torque component present (but undetectable) in all pulsars, \emph{plus} the sharp switching component seen only in a subset.  To investigate this hypothesis, we plot in Figure \ref{fig:s2-age} the rms timing noise $\sigma_2$, computed after fitting for $\nu$ and its first two time derivatives, as taken from Table 1 of \citet{hlk10}, versus characteristic age. 
As can be seen readily by eye, the subset do have somewhat larger amounts of timing noise than other pulsars of similar ages, but only by a factor of a few providing modest but by no means compelling evidence in favour of this model.

%%%%%%%%%%%%%%%%%%%%%%%%%%%%%%%%%%%%%%
\section{Discussion}  \label{sect:discussion}
%%%%%%%%%%%%%%%%%%%%%%%%%%%%%%%%%%%%%%

\subsection{Other possible explanations}

We have argued that the free precession hypothesis is not an alternative to the magnetospheric switching model advanced in \citet{letal10}; the two are compatible, with the precession phase-dependent accelerating potentials causing the switching from one magnetospheric state to the other.  However, several other models have been advanced to explain harmonic timing structure.  Most were developed before the switching behaviour was reported.  We comment on them briefly here.

\subsubsection{Tkachenko waves}

As was first pointed out by \citet{rude70_sf}, there is, in addition to free precession, another form of stellar oscillation that might give modulation periods of order a year: Tkachenko oscillations in a rotating superfluid.  These are oscillations in the superfluid vortex array.  The oscillations would result in a slight variation in the stellar moment of inertia.  By angular momentum conservation, this would manifest itself in timing variations.  The Tkachenko wave speed is given by $c_{\rm t}^2 = \kappa \Omega/8\pi$, where $\kappa$ is the quantum of vorticity given in terms of the neutron mass $m_{\rm n}$:  $\kappa =h/2m_{\rm n}$.  For an oscillation of wavelength $\lambda$ this leads to a mode period of order
\begin{equation}
\label{eq:P_T}
P_{\rm T} \approx 1.4 {\, \rm years \,} \left(\frac{P}{1 \, \rm s}\right)^{1/2}  \left(\frac{\lambda}{10^6 \, \rm cm}\right) .
\end{equation}
More detailed multi-fluid calculations  by \citet{hask11}  and \citet{ns08} confirm the existence of such long period oscillations (for some portion of the microphysical parameter space at least), leading these authors to suggest Tkachenko oscillations as an explanation for the observe timing variations, alternative to free precession.  Also of interest is the model advanced by \citet{popo08}, where Tkachenko oscillations couple to free precession in those stars where $P_{\rm fp} \approx P_{\rm T}$. 

The attractive feature of an explanation based on Tkachenko oscillations  lies in its lack of fine tuning---only basic properties of vortices go into equation (\ref{eq:P_T}).  However, there are a number of problems.  Firstly, the best-fit scaling between $P_{\rm mod}$ and $P$ is much closer to $P_{\rm mod} \propto P$ than to $P_{\rm mod} \propto P^{1/2}$, favouring precession over Tkachenko waves; see equation (\ref{eq:best_fit}).  Secondly, the long periodicities of \citet{hask11}  and \citet{ns08} are found by a plane wave analysis on a uniform background, and, crucially, for wave vectors exactly perpendicular to the vortex array.   For wave vectors not perpendicular to the rotation, the oscillation frequency is increased to a value of order the spin frequency.  In a real star a global mode must be calculated in a spheroidal-like geometry, with regularity conditions at the centre and boundary conditions at the spheroidal surface.  It is not clear if enforcing such a geometry will still allow for long-period oscillations.
 Finally, if the Tkachenko oscillations are to cause magnetospheric switching, presumably significant surface motions of the crust must occur, to create the time-varying surface electric fields that we have argued are needed.  Such crustal motions will provoke elastic restoring forces, again raising the mode frequency to too high a level.

\subsubsection{Slow glitches} \label{sect:sg}

Shabanova has attributed the interesting timing behaviour of three of the pulsars in our sample (PSRS B0919+06, B1642-03 and B1822-09) to the phenomenon of `slow glitches', where there is a slow rather than step-like increase in spin frequency, over a period of a few hundred days.  As noted by Shabanova, there are a number of  arguments in favour of the slow glitch interpretation over the precession one.  Firstly, in PSR B1642-03, the timing residuals are not quite periodic \citep{shab09}.  Instead the spacing between peaks seems to vary by a factor of about $2.6$ (but, interstingly, there seems to be some regularity in the this variation; see below).   Secondly, the residuals of PSR B0919+06 do have a clear periodic structure, but seem to undergo a $1/3$ of a cycle shift during the observing period \citep{shab10}.  Thirdly, in PSR B0919+06 there is no observed variation in pulse profile, as would be expected if precession were at work \citep{shab10}.

While attractive, the slow glitch  model has two drawbacks.  Firstly there is no obvious mechanism to provide magnetospheric switching, something which has been reported for PSR B1822-09 in \citet{letal10}.  Secondly, there is no obvious mechanism to provide the well-defined periodicity (modulo the reported phase change) seen in PSR B0919+16 or the approximately periodicity claimed for PSR B1642-03.  Given this, we can ask just how serious are the difficulties with the precession interpretation given above.  

As regards the timing issue, as noted in section \ref{sect:stability}, a good measure of the timing stability is the product $T_{\rm s}\Delta F$, as this averages over the duration span, minimising the effects of the non-precessional contributions to the timing noise.  The values of this quantity for the three pulsars are $1.1, 2.5, 1.4$, not significantly in excess of unity.  So, averaged over the full data span, these objects are stable in modulation frequency.  Also, for B1642-03, Shabnova notes that there is some regularity in the time intervals between peaks in the phase residuals, with a $60$ year overall envelope being tentatively identified \citep{shab09}.  That the residuals should have a more complex structure than a simple sinusoid is perfectly natural if the body has a significant degree of triaxiality, as then the timing is sensitive to both the $I_3-I_1$ and $I_2-I_1$ moment of inertia differences \citep{ll69, alw06}.  We therefore suggest that the timing data for the pulsars in question is not fatal for the precession interpretation of their behaviour.

Finally, what of the lack of pulse profile variation in PSR B0919+06?  As first pointed out by \citet{rude70}, precession with wobble angle $\theta$ should result in the observer's cut through the beam varying by an angle $\pm \theta$ once per precession cycle.  Why is this not apparent?  A possible explanation might lie in magnetospheric switching.  If switching occurs in this pulsar then, as explained above, the necessary wobble angle $\theta$ might then be very small.  If the component of the magnetospheric emission that is being switched on and off does not produce emission toward Earth, then only the smooth beam variation caused by the small precession angle would be apparent, which might be too small to be discernible.

\subsubsection{Asteroids and orbital companions}

Low mass binary  companions are a natural and simple way of explaining long-term periodicities in pulsar data, but  can a low-mass companion explain the magnetospheric switching phenomenon?  The binary companion model was analysed by \citet{retal08} for the intermittent pulsar PSR B1934+2.  The companion was modelled as a low mass star on an eccentric $\sim 35$ day orbit.  When at periastron the companion influences the pulsar emission, either by supplying additional plasma to power increased magnetospheric emission, pushing the pulsar from the off to the on state, or by causing transitory wind accretion onto the pulsar, pushing it into the propellor regime, i.e. from the on to the off state.  However, \citet{retal08} found that the orbit had to be extremely close to edge-on for this model to work, and that the likely wind from the low-mass companion was probably too low to affect the magnetosphere sufficiently.  Whether or not the lower level of magnetospheric switching that has been seen in other pulsars can be explained by planetary companions is not clear.

Regardless of this, the planetary hypothesis provides no good explanation for the distribution of modulation periods as a function of spin period of Figure \ref{fig:P_mod_v_P}: there is no reason to expect any particular relationship between the orbital and spin frequencies of companion and pulsar.  This last point also applies to the asteroid model of \citet{cs08}, where ionisation of an infalling asteroid was used to supply the changing magnetospheric activity.

\subsubsection{Non-radial modes}

Recently \citet{rmt11}  have proposed a model to explain the findings of \citet{letal10}. Their model has a significant overlap with the one proposed here, as they too suppose that  velocity variations in the surface of the star cause modulations in the magnetospheric spin-down torque. In their model precession is not the cause of this variation. Instead, the variations are due to the presence of non-radial oscillation modes with non-zero amplitudes at the stellar surface. The mode itself is of short period: the long-term variations are to be understood as being caused by variations in the amplitude of excitation of a single mode, or the redistribution of energy over several modes.   While attractive, their model does not (as far as we are aware) provide a natural explanation of why the mode excitation amplitudes would undergo the long-term fluctuations necessary to explain the long-term periodic structure seen in the timing residuals or for the approximately linear scaling of this modulation period with the spin period (as per Figure \ref{fig:P_mod_v_P}).

%%%%%%%%%%%%%%%%%%%%%%%%%%
\subsection{Further possible tests} \label{sect:fpt}

Having carried out a critical appraisal of the precession hypothesis, it is natural to ask what further tests might be carried out, that might strengthen or weaken the evidence in its favour.

The most obvious test would be to identify more precession candidates, particularly candidates with spin periods very different from those in the data set of \citet{letal10}.   The prospect of finding precession in longer spin period pulsars does not seem good. The \citet{letal10} pulsars had periods in the range $0.1 \lesssim P \lesssim 1.0$ second.  Assuming that $P_{\rm mod} \sim P / 5 \times 10^{-9}$, then a spin period $P \gtrsim1$ seconds implies a modulation period $P_{\rm mod} \gtrsim 6$ years, too long to be visible in all but the longest of data sets.  So, shorter spin period pulsars, with $P < 0.1$ seconds would provide more useful data. 

As was first pointed out by \cite{shah77}, the inclusion of a pinned superfluid component alters the relation between $P_{\rm mod}$ and $P$ drastically.  If we assume small wobble angle $\theta$ and that the moment of inertial difference $I_3-I_1$ is much less than the moment in inertia of the pinned superfluid, the $P_{\rm mod}(P)$ relation becomes
\begin{equation}
\label{eq:P_mod_pinning}
P_{\rm mod} = P \frac{I_{\rm C}}{I_{\rm SF}}  = P \frac{I_{\rm C}}{I_\star} \left(\frac{I_{\rm SF}}{I_\star}\right)^{-1},
\end{equation}
where $I_{\rm SF}$ and $I_{\rm C}$ denote the moments of inertia of the pinned superfluid and crust, respectively \citep{swc99, gaj09}.  Note that by `crust' we again mean the actual crust plus all other parts of the star coupled to it that participate in the precession.  The ratio of $P_{\rm mod}$ to $P$ then depends upon the extent of the pinning (i.e. on $I_{\rm SF}$) and on how much of the star participates in the precession (i.e. on $I_{\rm C}$).  There are several possibilities.  In the case of pinning only in the inner crust, we have $I_{\rm SF} \sim 10^{-2} I_\star$.  If the core neutron fluid does not participate in the precession, we have $I_{\rm C} \sim 10^{-2}$, giving $P_{\rm mod} \sim P$, while if the core neutron fluid does participate, we have $I_{\rm C} \sim I_\star$ giving $P_{\rm mod} \sim 10^2 P$.  However, if there is pinning in the core we have $I_{\rm SF} \sim I_\star$, so that if the core fluid does not participate (i.e. $I_{\rm C} \sim 10^{-2} I_\star$) we have $P_{\rm mod} \sim 10^{-2} P$, while if the core fluid does participate (i.e. $I_{\rm C} \sim I_\star$) we have $P_{\rm mod} \sim P$.

Clearly, the modulation periods are very short in all of these cases as compared to the no superfluid pinning case, and may even be shorter than the spin period.  It is not at all obvious how such a fast modulation would best be identified in pulsar data.  What is clear is that observation of any one of these cases would then mean that an alternative non-precessional explanation must be found for the periodicities discussed in \citet{letal10}.

\begin{figure} 
\caption{\label{fig:theory_P-P_mod} Some possible relations between modulation period $P_{\rm mod}$ and spin period $P$.  The line labelled ``Relaxed crust'' was obtained from equation (\ref{eq:P_mod_relaxed}) with $b = 5 \times 10^{-7}$.  The line labelled ``Stressed crust, no pinning'' was obtained from equations (\ref{eq:P_mod_fast}) and (\ref{eq:P_mod_slow}) with $b =  10^{-7}$, $u = 5 \times 10^{-4}$and $I_{\rm C}/I_\star = 10^{-2}$.  The line labelled ``Tkachenko'' was obtained from (\ref{eq:P_T}) with $\lambda = 10^6$ cm.  The line labelled ``Crustal pinning'' was obtained by putting $I_{\rm SF}/I_{\rm C} = 10^{-2}$ in (\ref{eq:P_mod_pinning}).  The line labelled ``Core pinning'' was obtained by putting $I_{\rm SF}/I_{\rm C} = 10^{2}$ in (\ref{eq:P_mod_pinning}).  } 
\includegraphics[angle=-90, width=90mm]{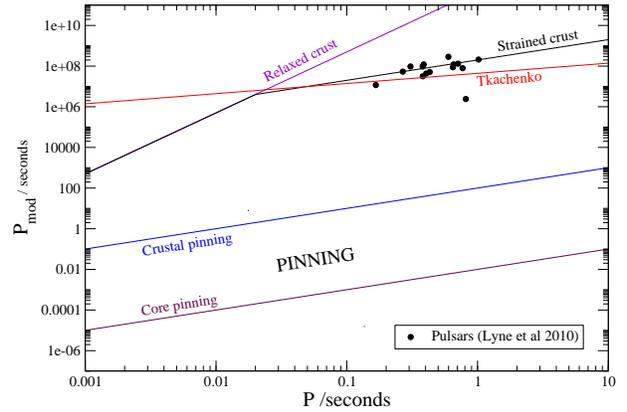} 
\end{figure}

To summarise these ideas, and help guide observers as to which pulsar populations they might wish to target, we plot several curves of possible $P_{\rm mod}(P)$ relations in Figure \ref{fig:theory_P-P_mod}.
The line for a star precessing with an elastically relaxed crust  corresponds to equation (\ref{eq:P_mod_fast}) with $b = 10^{-7}$ and $I_{\rm C}/I_\star = 10^{-2}$.  We also plot the precession period for a star which, at short rotation periods, can be approximated as having a relaxed crust, but at long rotation rates has a reference oblateness determined by the breaking strain $u_{\rm break}$.  These lines were obtained by setting $b = 10^{-7}$, $u = u_{\rm break} = 5 \times  10^{-4}$ and $I_{\rm C}/I_\star = 10^{-2}$ in  equations (\ref{eq:P_mod_fast}) and (\ref{eq:P_mod_slow}).  The critical spin period dividing the two regimes is $P_{\rm crit} \sim 20.5 (5 \times 10^{-4}/u_{\rm break})^{1/2}$ ms (see   equation (\ref{eq:P_crit})).
As noted previously, for realistic breaking strains, the current precession candidates all lie on the $P > P_{\rm crit}$ part of the curve where $P_{\rm mod}$ is a linear function of $P$, with $P / P_{\rm mod} \sim b u_{\rm break} (I_\star/I_{\rm C})$ , so that observations of $P_{\rm mod}$ and $P$ only allow the extraction of the product of these factors.   Data points corresponding to pulsars in the higher spin rate relaxed regime, where $P_{\rm mod}$ scales as $P^3$, would allow the identification of the location of the knee, which  would allow the value of $u_{\rm break}$ to be extracted separately.  Clearly, observations of more periodicities in more rapidly spinning pulsars would be needed to do this, and if $u_{\rm break}$ is as large as the recent microphysical modelling indicates \citep{hk09}, then few if any pulsars will reside in this regime.  Setting the moment of inertia ratio $I_{\rm C}/I_\star$ equal to $1$ rather than $10^{-2}$ would simply move these curves downwards to be a factor of $10^2$.
The line for Tkachenko oscillations is given by equation (\ref{eq:P_T}).  Its similar slope to the previously described line again suggests that observations of spin-down in milli-second pulsars (MSPs) would be very useful.   However, the most striking feature of the plot is the well known many order-of-magnitude difference between the pinning and no-pinning predictions.  Alternatively, for the precession of stars deformed by superconducting fields, Figure \ref{fig:P_mod_v_age} is the relevant one.  In this case the shortest modulation periods would be $\sim 0.1$ years, for the youngest pulsars, while the longest modulation periods would be of order hundreds of years, for the MSPs.  Clearly, observers will need to be allow for modulation over a wide range of timescales to cover all of the theoretical predictions.

In addition to the use of pulse profiles and timing residuals, pulse polarisation is also a useful quantity to measure when looking for precession.  According to the rotating vector model \citep{rc69}, because the pulse polarisation is tied to the magnetic field, which is in turn tied to the stellar crust, the precession would manifest itself as a longer-period  variation superimposed on the polarisation angle  sweeps that occur once per rotation.  Indeed, a recent study of polarisation measurements was carried out by \citet{wetal10}, who found that the position angle variations in $19$ pulsars were significantly better fit by a sinusoid than by a constant.  However, we have not carried out a statistical fit to this data as only one of the the $4$ pulsars in the `Class I' i.e. the subset of pulsars considered most likely by the authors to display sinusoidal variations, meets our selection criterion of $N>3$.  Clearly, analysis based on observations on a longer time baseline would be very interesting.

It would be particularly interesting to see if the sharp change in pulse profile observed in some pulsars is also accompanied by a sharp change in polarisation angle.  It may be the case that the overall magnetic field structure is not significantly affected by whether or not the magnetospheric process is on or off, but rather simply acts as a guide, determining the motion of the charged particles when the process is active.  In this case, the polarisation angle may undergo perfectly smooth variations, tied to the underlying precessional motion, despite the sharp switches in pulse profile.  However, it may be the case that the magnetospheric process, by switching on magnetospheric currents, does change the gross magnetic field structure.  In this case sharp switches in polarisation would occur.  Clearly there is some interest in looking at the polarisation variations in stars with sharp magnetospheric switches, but the presence of a sharp change in polarisation does not necessarily invalidate the precessional model proposed here.

%%%%%%%%%%%%%%%%%%%%%%%%%%%%%%%%%%%%%%%%%
\subsection{Comparison with gravitational wave spin-down upper limits}

In our simple biaxial precession model, an important quantity in determining the modulation period was the $I_3-I_1$ asymmetry in the moment of inertia tensor; we defined $\epsilon_{31} = (I_3-I_1)/I_\star$, and this is related to the observable  $P / P_{\rm mod}$ via equation (\ref{eq:P_over_P_fp}), so that
\begin{equation}
\label{eq:epsilon_31_histogram}
\epsilon_{31} = \frac{P}{P_{\rm mod}} \frac{I_{\rm C}}{I_\star}
\end{equation}
As was noted previously, a real star will be triaxial, not biaxial, which would lead to a  more complicated  free precessional motion \citep{alw06}.  Nevertheless, the observed values of the ratio $P/P_{\rm mod}$ are giving us information on the deformation of the stars' moment of inertia tensors, modulo the uncertainty in the value of the ratio $I_{\rm C}/I_\star$.

There is another aspect of pulsar physics that probes asymmetries in the moment on inertia tensor: gravitational wave emission.  Specifically, the observed spin-down rates of non-precessing stars can be used to place upper limits on the $I_2-I_1$ part of the inertia tensor.  If we define 
\begin{equation}
\epsilon_{21} = \frac{I_2-I_1}{I_\star} ,
\end{equation}
then an upper limit on this quantity is obtained by assuming that all of the kinetic energy loss associated with a spin-down rate $\dot P$ for a non-precessing pulsar with spin period $P$ is going into gravitational wave energy emission, leading to the upper limit
\begin{equation}
\epsilon_{21}^{\rm UL} = \left[ \frac{5c^5 \dot P P^3}{32(2\pi)^4 GI_\star} \right]^{1/2} ,
\end{equation}
(see e.g. \citet{aetal07}). This is an upper limit because in reality some portion of the spin down, possibly most, is accounted for by electromagnetic rather than gravitational wave emission.  

\begin{figure} 
\caption{\label{fig:e_hist} Histogram of estimated ellipticities $\epsilon_{31}$ derived from precession candidates and, for comparison, histogram of $\epsilon_{21}^{\rm UL}$ upper limits derived assuming $100\%$ gravitational wave spin-down for the known pulsar population.   Two versions of the $\epsilon_{31}$ histogram are given: the filled one has $I_{\rm C}/I_\star = 10^{-2}$, while the shaded one has $I_{\rm C}/I_\star = 1$.  The plot is truncated at ellipticities of $10^{-4}$ so as to restrict attention to the physically plausible values of $\epsilon_{21}^{\rm UL}$ corresponding to members of  the MSP population.  The non-MSP have unphysically large values of $\epsilon_{21}^{\rm UL}$; the beginning of this distribution can be seen clearly in the rightmost bin.   The outlier in the two $\epsilon_{31}$ distributions is PSR B1931+24.} 
\includegraphics[angle=-90, width=90mm]{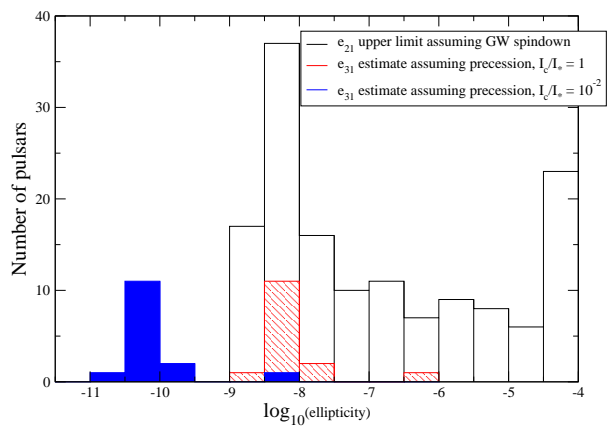} 
\end{figure}

The ellipticity $\epsilon_{21}$ can be non-zero only by virtue of strains within the star, so it is plausible that the mechanisms that limit the value of $\epsilon_{31}$ (as discussed in section \ref{sect:etmp}) will also limit the ellipticity $\epsilon_{21}$.  Given this, it  is interesting to compare the estimates of $\epsilon_{31}$ from precession, with the upper limits on $\epsilon_{21}$ obtained from spin-down.  A histogram of these quantities is given in Figure \ref{fig:e_hist}.
We plot the histogram only for ellipticities less that $10^{-4}$, for which the $\epsilon_{21}^{\rm UL}$ values correspond to members of the MSP population and are relatively plausible.  (The non-MSP population part of the histogram consists of a second broad peak, extending all the way up to ellipticities  of order unity, with a maximum  at ellipticities of around $10^{-2}$. This is a consequence of the well known fact that the $\epsilon_{21}^{\rm UL}$ upper limit ellipticities for the non-MSP population are unphysically  large.). 

We plot two versions of the $\epsilon_{31}$ histogram.  In one we set $I_{\rm C}/I_\star = 10^{-2}$ in equation (\ref{eq:epsilon_31_histogram}); this peaks around values of a few times $10^{-11}$.  In the other we set $I_{\rm C}/I_\star = 1$ in equation (\ref{eq:epsilon_31_histogram}); this peaks around values of a few times $10^{-9}$.  Amusingly,  this coincides with the peak in the histogram of upper limits on $\epsilon_{21}^{\rm UL}$ for the MSP population.   On the basis of this seeming coincidence,  one might speculate that $I_{\rm C}/I_\star$ really is close to unity and the physical mechanism limiting the two sorts of deformation is indeed one and the same.  Given our analysis of section \ref{sect:etmp} for the precession candidates we would then ascribe this to a highly strained neutron star crust, so that the product $b u_{\rm break}$ really is of order a few times $10^{-9}$, and that stars really do retain this level of \emph{non-axisymmetric} deformation, so that the upper limits on $\epsilon_{21}$ of the millisecond pulsar population are in fact estimates, not upper limits.  If true, this would clearly be of interest for gravitational wave astronomers, and would add to the motivation of continuing to target the MSPs in the advanced detector era.   That neutron star crusts might be strong enough to support such levels of stress was demonstrated previously by \citet{ucb00} and \citet{cul03}; all that is new here is the additional evidence in support of this, in the form of the histogram of $\epsilon_{31}$ values supplied by the multiple precession candidates.
 
If it were to be the case that the spin-down upper limit ellipticities $\epsilon_{21}^{\rm UL}$ of the  MSPs are indeed close to the actual ellipticities, it would follow that electromagnetic energy losses are smaller then currently believed, which in turn would imply that the MSP magnetic field strengths are lower than the $\sim 10^9$ G inferred on the basis of magnetic dipole spin-down.  Of course, the very fact that MSPs are visible as pulsars at all implies that their magnetic fields are not zero, as some level of magnetic field is required to power the electromagnetic pulsations.  Furthermore, the X-ray and gamma-ray emission from MSPs is correlated with the total spin-down luminosity \citep{petal02, ctw06}.  This implies that whatever is causing the spin-down is also playing a role in producing the high energy electromagnetic emission, as would be expected for magnetic dipole spin-down, but not for gravitational wave spin-down from a star whose non-axisymmetry is supplied by elastic strain.  This argues against a significant gravitational wave component to the spin-down.  We simply note that the coincidence between the peaks in the ellipticity histograms in the $I_{\rm C}/I_\star = 1$ case provides some small  evidence that neutron star crusts might be strong enough to support the level of deformation needed to power gravitational wave emission from MSPs.

%%%%%%%%%%%%%%%%%%%%%%%%%%%%%%%%%%%%%%%%%%%%
\section{Summary and conclusion} \label{sect:summary}
%%%%%%%%%%%%%%%%%%%%%%%%%%%%%%%%%%%%%%%%%%%%

We have presented a model of pulsar timing where the harmonic structure of timing residuals seen in \citet{letal10} is caused by free precession.  We argued that the relationship between the modulation period and spin period was of the correct form to be explained by free precession of an elastically strained star, the strain being due to the star retaining  a memory of a former faster spin rate \citep{cul03}.  The details depend upon the fraction of the star that participates in the precession.  In the case where only the crust participates ($I_{\rm C}/I_\star \approx 10^{-2}$) we found the level of strain to be $u \approx 5 \times 10^{-4}$, and the stars must have been born spinning as least as rapidly as $\approx 50$ Hz.  In the case where the whole star participates in the precession  ($I_{\rm C}/I_\star \approx 1$) we found the level of strain to be $u \approx 5 \times 10^{-2}$, and the stars must have been born spinning as least as rapidly as $\approx 500$ Hz.  In either case, the roughly identical levels of strain in the different pulsars might then reflect the universal breaking strain that a crust can withstand.  Alternatively, the nearly identical strain levels may reflect nearly identical spin rates at birth. 

We argued that the sharp switching in pulse profile that occurs in at least some of the candidate precessors  is a consequence of the star being delicately balanced between two different magnetospheric states,  with one state being statistically favoured over the other depending upon the precession phase.  That the link between magnetospheric state and precessional phase in only statistical was, we argued, consistent with the completely random mode-switching seen in some non-precessing stars.  The sharpness of the switching reflects the energy of particles at some point in the magnetosphere exceeding some critical threshold for an avalanche-like process, probably pair-production.

Within this model, the magnetosphere variations, locked in phase with the precession, modulate the spin-down torque,  serving to make free precession more readily detectable in such delicately balanced stars as compared to the rest of the population.  It is therefore no surprise that magnetospheric switching has been seen in precession candidates.  The need for the star to be delicately balanced between two different magnetospheric states may play a role in explaining the rarity of the precession phenomenon in the pulsar population, although the precession excitation mechanism (perhaps glitching) may play a role too.  

We noted in passing that, in the case where the whole star participates in the precession,  the required asymmetries in the moment of inertia tensor to explain the precession ($\epsilon_{31} \sim 5 \times 10^{-9}$) are similar to the upper limits on asymmetry $\epsilon_{21}$ that come from assuming the gravitational wave driven spin-down of the millisecond pulsar population.   This may offer hope to gravitational wave astronomers that these upper limits may in fact be estimates, making the targeting of the MSPs by future advanced detectors all the more interesting.  However, if only the crust participated in the precession, the corresponding estimates of $\epsilon_{31}$ are reduced by two orders of magnitude, comfortably below the upper limits on $\epsilon_{21}$, removing this seeming coincidence.

The analysis of crustal strain in precession candidates presented here improves upon that of previous works \citep{ja01, cul03} by considering a larger set of pulsars, as supplied by \citet{letal10}, and also by imposing the quality cut that at least three precession cycles should fit into the data set for the precession claim to be taken seriously; we suggest that future investigators use a similar quality cut.  Also, the analysis here improves on that of \citet{ja01} by using the accurate value of the rigidity parameter $b$ calculated by \citet{cul03}.

There are a number of objections to the model presented here.  Much of our argument is qualitative, rather than quantitative.  The only quantitative piece of evidence we have is the expected value of the modulation period for the slowly spinning stars discussed here, and its scaling with spin frequency, as illustrated in Figure \ref{fig:P_mod_v_P}.  However, the spread in spin periods (and modulation periods) was not large; more data points are needed to make this convincing.  

There is also a serious theoretical problem.  As pointed out by \citet{ja01} and \citet{le01}, the superfluid pinning that is believed to be responsible for the larger Vela-like pulsar glitches would drastically reduce the modulation period, eliminating precession as a possible explanation of the observed periodic timing data.  As has been argued by \citet{link03, link06}, if precession is indeed the mechanism responsible for the timing variations, it would be necessary for any superfluid parts of the star to be completely separated from the charged component, or at least that component that participates in the precession.  This conflicts with microphysical modelling of neutron star interiors, which indicates that the interior superfluid is likely to coexist with a sold crust and, deeper within the star, with magnetic flux tubes.

There is a caveat to be attached to this argument.  The argument against the starquake model of Vela-like glitches is as follows.  The strain relieved at each glitch is much larger than the strain build up by spin-down between glitches, so that the strain cannot possibly be replenished between glitches, i.e. a steady-state solution does not exist.  However,  the large crustal breaking strains  suggested by recent molecular dynamics solutions \citep{hk09} suggest a possible loop-hole.  A typical Vela glitch releases a strain of order $\Delta \nu / \nu \sim 10^{-6}$.  The Vela glitches approximately once every $5$ or so years, and is approximately $10^4$ years old.  This corresponds to a total of $\sim 10^3$ glitches over its lifetime, releasing a total strain of $10^{-3}$, two orders of magnitude smaller that the estimated breaking strain.  So, the Vela and similar pulsars may simply be in the process of divesting themselves of crustal strain built up by spinning down from a much faster rotation rate that applied earlier in their lives.  It follows that providing one is willing    for the Vela and similar glitching pulsars to be in such a non-steady state situation, their large glitches can have a non-superfluid explanation.  However, we can see no good reason why only a small fraction of the total strain should be relieved in each cracking event.

To sum up, we feel that the precession model advanced here is plausible, and gives a sensible account of the observed timing data.  However, the relatively modest number of pulsars with observations of periodic variation in spin-down, and their rather small spread in spin frequency, make forming a firm conclusion impossible.   A larger sample of pulsars with harmonic timing residuals is needed, spanning a wider range of spin and modulation periods.  By placing more points on the modulation period verses spin period plot (Figure \ref{fig:theory_P-P_mod}),  it should be possible to discriminate between the precession-based model proposed here and the handful of others that have appeared in the literature.  Also, a better picture of the stability of the clock that lies behind the timing data would be invaluable: modulo any irreversible changes in stellar shape, precession should provide high frequency stability; a clear demonstration of a lack of modulation stability would argue against the precessional hypothesis.

%%%%%%%%%%%%%%%%%%%%%%%%%%%%%%%%%%%%%%%%%%%%%%%%%%%%

\section*{Acknowledgments}

It is a pleasure to thank Kostas Glampedakis, Wynn Ho and David Kaplan for useful conversations during the course of this work.
The author acknowledges support from STFC via grant number ST/H002359/1, and also travel support from COMPSTAR (an ESF Research Networking Programme).

%%%%%%%%%%%%%%%%%%%%%%%%%%%%%%%%%%%%%%%%%%%%%%%%%%%

\bibliography{references}

\end{document}